\documentclass[pra,twocolumn,superscriptaddress,showpacs]{revtex4}
\usepackage{graphicx}

\begin{document}
\title{Lieb-Liniger model of a dissipation-induced Tonks-Girardeau gas}
\author{S. D\"{u}rr}
\affiliation{Max-Planck-Institut f{\"u}r Quantenoptik, Hans-Kopfermann-Stra{\ss}e 1, 85748 Garching, Germany}
\author{J.~J. Garc\'{i}a-Ripoll}
\affiliation{Max-Planck-Institut f{\"u}r Quantenoptik, Hans-Kopfermann-Stra{\ss}e 1, 85748 Garching, Germany}
\affiliation{Facultad de F\'{i}sicas, Universidad Complutense, Ciudad Universitaria s/n, Madrid 28040, Spain}
\author{N. Syassen}
\author{D.~M. Bauer}
\author{M. Lettner}
\author{J.~I. Cirac}
\author{G. Rempe}
\affiliation{Max-Planck-Institut f{\"u}r Quantenoptik, Hans-Kopfermann-Stra{\ss}e 1, 85748 Garching, Germany}

\begin{abstract}
We show that strong inelastic interactions between bosons in one dimension create a Tonks-Girardeau gas, much as in the case of elastic interactions. We derive a Markovian master equation that describes the loss caused by the inelastic collisions. This yields a loss rate equation and a dissipative Lieb-Liniger model for short times. We obtain an analytic expression for the pair correlation function in the limit of strong dissipation. Numerical calculations show how a diverging dissipation strength leads to a vanishing of the actual loss rate and renders an additional elastic part of the interaction irrelevant.
\end{abstract}


\maketitle

\section{Introduction}
A Tonks-Girardeau gas \cite{tonks:36,girardeau:60} is a one-dimensional (1D) system of identical bosons with strong interparticle interactions. The wave functions of the bosons have a surprisingly simple one-to-one mapping to a system of non-interacting fermions \cite{girardeau:60}. In recent years, a Tonks-Girardeau gas was realized in experiments with ultracold gases \cite{paredes:04,kinoshita:04}. A characteristic property of the Tonks-Girardeau gas is that the probability to find two particles at the same position is very small \cite{gangardt:03,kinoshita:05}. Theoretical and experimental studies of the Tonks-Girardeau gas so far dealt only with the case of elastic interactions. In a recent experiment, we studied a 1D gas in which the bosons strongly interact inelastically, leading to loss of particles \cite{syassen:08}. This system also shows a strong suppression of the probability to find two particles at the same position and, indeed, demonstrates an alternative way of realizing a Tonks-Girardeau gas. The broader impact of this experiment lies in the demonstration that dissipation can be used to drive a system into the strongly-correlated regime.

Here we study the theory of the dissipative Tonks-Girardeau gas. We show that in the limit of infinite interaction strength Girardeau's solution \cite{girardeau:60} is reached, as in the case of strong elastic interactions. We derive a Markovian master equation that describes the dissipation in the system and yields a rate equation for the particle loss. When the study is restricted to loss rates at short times, the master equation can be reduced to a Schr\"{o}dinger equation with a non-Hermitian version of the Lieb-Liniger Hamiltonian \cite{lieb:63}. This approach yields an analytic expression for the loss rate in the limit of strong dissipation. Furthermore, we present numerical solutions to the dissipative Lieb-Liniger model at finite interaction strength that illustrate how a possible elastic part of the interactions becomes irrelevant as the dissipation strength diverges. In addition these numerical calculations reveal the dependence of the pair correlation function on dissipation strength, showing that an increase in the dissipation strength leads to a decrease of the actual loss rate due to the build-up of strong correlations. Finally, we present an estimate for the loss rate at longer times.

\section{Inelastic Ultracold Collisions}
In this section, we discuss a master equation approach to describe loss processes caused by inelastic two-body collisions in an ultracold gas. First, we introduce the master equation, then use it to derive a rate equation for the particle loss, and finally show that in the short-time limit, a non-Hermitian Lieb-Liniger model is obtained.

\subsection{Master Equation}
We consider a three-dimensional (3D), dilute, ultracold gas of identical bosons, which all occupy the same internal state. Two-body collisions in such a gas can be described in terms of the 3D $s$-wave scattering length $a$. For elastic collisions, it is customary to replace the interparticle scattering potential by a delta potential $g_{\rm 3D} \delta^{(3)}({\bf x})$ with interaction strength \cite{huang:87Sec10.5}
\begin{eqnarray}
\label{g-3D}
g_{\rm 3D} = \frac{4\pi\hbar^2 a}m
,
\end{eqnarray}
where $m$ is the mass of a particle. For the many-body system in second quantization, this yields a Hamiltonian (see appendix \ref{app-master} or, {\it e.g.}, Ref.\ \cite{pitaevskii:03})
\begin{eqnarray}
\label{H-0}
H_0 &=& \int d^3x \Psi^\dagger ({\bf x}) H_s({\bf x}) \Psi ({\bf x}) 
\nonumber \\ &&
+ \frac{{\rm Re}(g_{\rm 3D})}{2} \int d^3x \Psi^{\dagger2}({\bf x}) \Psi^2 ({\bf x}) 
,
\end{eqnarray}
where $H_s({\bf x})=-\hbar^2\nabla^2/2m$ is the single-particle Hamiltonian in first quantization and $\Psi({\bf x})$ is the field operator that annihilates a boson at position $\bf x$. In the case of purely elastic interactions $g_{\rm 3D}$ is real anyway, so taking the real part of it in Eq.\ (\ref{H-0}) has no effect.

We now generalize this treatment to include inelastic collisions. First, we consider inelastic collisions on the level of a two-body problem. This yields an imaginary part of the scattering length, with ${\rm Im}(a)\leq 0$ \cite{mott:65,bohn:97,balakrishnan:97,hutson:07}. This is because the elastically scattered wave is proportional to the diagonal element of the $S$-matrix $e^{-2iak_{\rm rel}}$, where ${\bf k}_{\rm rel}$ is the wave vector in the relative motion. Hence, a negative imaginary part of $a$ reduces the elastically scattered flux. This missing flux appears in the inelastically scattered channels.

Next, we consider the effect of inelastic collisions on the level of the many-body system. The interaction strength $g_{\rm 3D}$ defined by Eq.\ (\ref{g-3D}) becomes complex-valued. Equation (\ref{H-0}) contains only the real part of $g_{\rm 3D}$ and thus represents only the elastic part of the interactions (see appendix \ref{app-master}). For a treatment of the inelastic collisions, it is crucial what happens to the collision partners after the collision. The difference in internal energy that is released in the change of the internal state appears in the kinetic energy of the relative motion of the particles after the collision. In typical experiments with trapped ultracold gases, the energy released here is so large that all particles involved in the collision quickly escape from the trap. This loss of particles is an irreversible process. In appendix \ref{app-master}, we derive a Markovian quantum master equation
\begin{eqnarray}
\label{master}
\hbar \frac{d\rho}{dt} &=& - i [H_0,\rho] + {\cal D}(\rho)
\end{eqnarray}
for the time evolution of the density matrix $\rho$ of this many-body system, with a dissipator
\begin{eqnarray}
\label{D}
{\cal D}(\rho) &=& - \frac{{\rm Im}(g_{\rm 3D})}{2} \int d^3x 
\left( 2 \Psi^2({\bf x}) \rho \Psi^{\dagger 2}({\bf x}) \right. 
\nonumber \\ &&
\left. 
- \Psi^{\dagger 2} ({\bf x}) \Psi^2 ({\bf x}) \rho 
- \rho \Psi^{\dagger 2}({\bf x})  \Psi^2 ({\bf x}) 
\right)
\end{eqnarray}
that has a Lindblad form \cite{carmichael:99,breuer:02}.

The dissipator describes loss of pairs of particles due to inelastic collisions. More specifically, the last two terms in $\cal D$ deplete terms in the density matrix that represent states with more than one particle. The first term in $\cal D$ makes this lost population reappear in states with two fewer particles. This is further illustrated in Sec.\ \ref{sec-short-times}.

\subsection{Loss Rate in 3D}
\label{sec-loss-3D}
We now use this master equation approach to derive a rate equation for the loss of particle number. The particle density operator $\hat n_{\rm 3D}({\bf x}) = \Psi^\dagger({\bf x}) \Psi({\bf x})$ has an expectation value that has a time dependence with a contribution from $H_0$, which conserves the total particle number, and a contribution from ${\cal D}(\rho)$, which describes loss
\begin{eqnarray}
\left. \frac{d\langle \hat n_{\rm 3D}({\bf x})\rangle}{dt} \right|_{\rm loss}
&=& \frac1\hbar {\rm Tr}[\hat n_{\rm 3D}({\bf x}) {\cal D}(\rho)]
\\
\label{3D-loss-temp}
&=& -K_{\rm 3D} \langle \Psi^{\dagger 2}({\bf x}) \Psi^2({\bf x})\rangle
,
\end{eqnarray}
where Tr denotes the trace and
\begin{eqnarray}
\label{K-3D}
K_{\rm 3D} = - \frac2\hbar {\rm Im}(g_{\rm 3D}) = -\; \frac{8\pi\hbar}m {\rm Im}(a)
\end{eqnarray}
is the rate coefficient for two-body loss. The loss-rate equation (\ref{3D-loss-temp}) can be rewritten as
\begin{eqnarray}
\label{dndt-3D}
\left. \frac{d\langle \hat n_{\rm 3D}({\bf x})\rangle}{dt} \right|_{\rm loss} = - K_{\rm 3D} \langle \hat n_{\rm 3D}({\bf x})\rangle^2 g^{(2)}({\bf x})
\end{eqnarray}
with the pair correlation function
\begin{eqnarray}
\label{g2-def}
g^{(2)}({\bf x})
= \frac{\langle \Psi^{\dagger2}({\bf x}) \Psi^2({\bf x}) \rangle}{\langle \hat n_{\rm 3D}({\bf x}) \rangle^2}
.
\end{eqnarray}
If $N$ particles occupy the same single-particle wave function, such as in a Bose-Einstein condensate (BEC), one obtains $g^{(2)}=(N-1)/N$. For $N\rightarrow\infty$ such a system is uncorrelated, {\it i.e.}, $g^{(2)}=1$. For comparison, a thermal gas of bosons above the critical temperature for Bose-Einstein condensation shows thermal bunching with $g^{(2)}=2$. $g^{(2)}$ quantifies how far the loss rate deviates from that of an uncorrelated system. Similar considerations apply to inelastic three-body collisions \cite{kagan:85,burt:97,laburthe:04}.

\subsection{Loss Rate in 1D}
\label{sec-loss-1D}
The above master-equation approach, has a straightforward generalization to a 1D system, yielding
\begin{eqnarray}
\label{dndt-1D}
\left. \frac{d\langle \hat n_{\rm 1D}(x)\rangle}{dt} \right|_{\rm loss} 
&=& - K_{\rm 1D} \langle \hat n_{\rm 1D}(x)\rangle^2 g^{(2)}(x)
\\
\label{K-1D}
K_{\rm 1D} &=& - \frac2\hbar {\rm Im}(g_{\rm 1D})
,
\end{eqnarray}
where $n_{\rm 1D}$ is the 1D particle density and $g_{\rm 1D}$ is the strength of the 1D delta potential, which can be related to the 3D scattering length $a$ as described now.

Experimental realizations of 1D systems typically use a strong harmonic confinement in the two transverse dimensions, with trap angular frequency $\omega_\perp$. If one approximates the 3D particle density along the two transverse dimensions by the ground state of the harmonic oscillator with oscillator length $a_\perp=\sqrt{\hbar/m\omega_\perp}$, then spatial integration of Eq.\ (\ref{dndt-3D}) over the two transverse dimensions yields the estimate
\begin{eqnarray}
\label{K1D-approx}
K_{\rm 1D} \approx \frac{K_{\rm 3D}}{2\pi a_\perp^2}
.
\end{eqnarray}

A more rigorous approach to this scattering problem in a confined geometry was presented in Ref.\ \cite{olshanii:98}, yielding
\begin{eqnarray}
\label{g-1D}
g_{\rm 1D} = \frac{2\hbar^2 a}{m a_\perp^2} \left[ 1+ \frac{a}{\sqrt2 \; a_\perp} \; \zeta\left(\frac12\right) \right]^{-1}
,
\end{eqnarray}
where $\zeta$ denotes the Riemann zeta function with $\zeta(\frac12)\approx -1.46$. In the limit $|a|\ll a_\perp$, this reproduces Eq.\ (\ref{K1D-approx}). But outside this regime, the scattering process is altered by the transverse confinement and Eq.\ (\ref{K1D-approx}) is no longer valid. The derivation in Ref.\ \cite{olshanii:98} considered only real-valued $a$, but following this derivation, one can show that Eq.\ (\ref{g-1D}) remains valid for complex-valued $a$. The applicability of Eq.\ (\ref{g-1D}) for the parameters of our experiment in Ref.\ \cite{syassen:08} is discussed in appendix \ref{app-experiment}.

\subsection{Loss Rate at Short Times}
\label{sec-short-times}

For studies of the behavior of the system at short times, it is useful to rewrite the master equation (\ref{master}) as
\begin{eqnarray}
\label{master-eff}
\hbar \frac{d\rho}{dt} &=& - i H_{\rm eff} \rho + i \rho H_{\rm eff}^\dagger + {\cal J}(\rho) 
\end{eqnarray}
with an effective Hamiltonian, that is not Hermitian and turns out to be the analytic continuation of $H_0$
\begin{eqnarray}
\label{H-eff}
H_{\rm eff} 
&=& H_0 + i \frac{{\rm Im}(g_{\rm 3D})}{2} \int d^3x \Psi^{\dagger2}({\bf x}) \Psi^2 ({\bf x}) 
\end{eqnarray}
and with
\begin{eqnarray}
{\cal J}(\rho) &=& - \; {\rm Im}(g_{\rm 3D}) \int d^3x \Psi^2({\bf x}) \rho \Psi^{\dagger 2}({\bf x})
.
\end{eqnarray}
Note that in a quantum Monte-Carlo approach to open quantum systems, $\cal J$ would be represented by the quantum jump part of the time evolution \cite{scully:99}.

The effect of the different terms in the master equation (\ref{master-eff}) can be illustrated by considering a density matrix that initially represents a homogeneous BEC with exactly $N$ particles. For short times Eq.\ (\ref{master-eff}) yields
\begin{eqnarray}
\label{xi}
\rho = (1-\xi t) |N\rangle\langle N| + \xi t |N-2\rangle\langle N-2| + {\cal O}(\xi^2 t^2)
,
\end{eqnarray}
where $\xi=K_{3D} n_{\rm 3D} (N-1)/2$. The loss rate of particle number is 
\begin{eqnarray}
\label{dNdt-xi}
\frac{d N}{dt} = -2 \xi [1+{\cal O}(\xi t)]
.
\end{eqnarray}

A detailed look at the calculation shows that $H_{\rm eff}$ causes the decay of the probability to find $N$ particles, whereas $\cal J$ causes the build-up of the probability to find $N-2$ particles. If we are interested only in calculating $dN/dt$ at short times, $t\ll 1/\xi$, we can use this observation to drastically simplify the model. The key idea is that refilling population into states with $N-2$ particles is important only when the evolution at long times is considered, namely at times where states with $N-2$ particles also decay significantly. But at short times, all that counts is how fast population is lost from the initial state.

Guided by this idea, we drastically simplify the model by dropping $\cal J$ from the model. This removes the term $\xi t |N-2\rangle\langle N-2|$ from Eq.\ (\ref{xi}). As a result, ${\rm Tr}(\rho)$ decays as a function of time, which is unphysical, but the rate at which this decay occurs 
\begin{eqnarray}
\label{dTrRhodt-xi}
\frac{d{\rm Tr}(\rho)}{dt} = -\xi [1+{\cal O}(\xi t)]
\end{eqnarray}
is quite informative, because this is the rate at which population is lost from the initial state. We know that in reality this population reappears in states with $N-2$ particles. As each loss removes two particles, the true loss rate of particle number can be estimated from the decay rate of ${\rm Tr}(\rho)$ in the simplified model. Indeed, comparison of Eqs.\ (\ref{dNdt-xi}) and (\ref{dTrRhodt-xi}) yields
\begin{eqnarray}
\label{dNdt-Tr-rho}
\left. \frac{d N}{dt} \right|_{t=0} = 2 \left. \frac{d{\rm Tr}(\rho)}{dt} \right|_{t=0}
.
\end{eqnarray}
Note that here $dN/dt$ refers to the true loss rate of particle number in the full model, whereas $d{\rm Tr}(\rho)/dt$ is a quantity calculated in the simplified model. The factor of 2 in this equation reflects the fact that two particles are lost per inelastic collision.

With $\rho$ initially prepared in a pure state $\rho=|\psi\rangle\langle\psi|$ and with $\cal J$ neglected, the master equation reduces to a Schr\"odinger equation with a non-Hermitian Hamiltonian $H_{\rm eff}$. According to Eq.\ (\ref{dNdt-Tr-rho}) the particle loss rate becomes
\begin{eqnarray}
\left. \frac{d N}{dt} \right|_{t=0} = \frac2{i\hbar} \left\langle H_{\rm eff} - H_{\rm eff}^\dagger \right\rangle
.
\end{eqnarray}
For a right eigenvector of the effective Hamiltonian $H_{\rm eff}|\psi\rangle=E|\psi\rangle$, we obtain
\begin{eqnarray}
\label{dN-dt-Im-E}
\left. \frac{d N}{dt} \right|_{t=0} = \frac4\hbar {\rm Im} (E)
.
\end{eqnarray}
The calculation of the initial loss rate is thus possible using a Schr\"odinger equation with $H_{\rm eff}$, instead of the master equation. The imaginary part of an energy eigenvalue $E$ is always non-positive and represents the initial loss rate of the particle number.

This greatly simplifies the calculation, because it suffices to study the eigenvectors and eigenvalues of an effective Hamiltonian, instead of performing a time-resolved calculation of the density matrix. We will follow this approach throughout Sec.\ \ref{sec-Lieb}.

\section{Dissipative Lieb-Liniger Model}
\label{sec-Lieb}
Motivated by the results of the previous section, we now study the eigenvectors and eigenvalues of the effective Hamiltonian. This effective Hamiltonian is a non-Hermitian version of the Lieb-Liniger model \cite{lieb:63}. We are mostly interested in the regime of strong dissipation. First, we show that in this regime, a Tonks-Girardeau gas is reached, second, we derive an analytical expression for the pair-correlation function in this regime and, third, we compare this with results of a numerical calculation, which show that a finite real part of the scattering length becomes irrelevant for diverging dissipation strength.

\subsection{Tonks-Girardeau Gas for Infinite Interaction Strength}
We consider a 1D gas of $N$ identical bosons in a box of length $L$ with periodic boundary conditions and with the 1D version of the effective Hamiltonian of Eq.\ (\ref{H-eff}). In first quantization this effective Hamiltonian reads
\begin{eqnarray}
\label{H-LL}
H_{\rm eff} = - \frac{\hbar^2}{2m} \sum_i \frac{\partial^2}{\partial x_i^2} + g_{\rm 1D} \sum_{i<j} \delta (x_i-x_j)
,
\end{eqnarray}
where $x_i$ is the position of the $i$th boson. For the case of real-valued $a$, this model was first considered by Lieb and Liniger \cite{lieb:63}, who introduced the dimensionless interaction strength
\begin{eqnarray}
\label{gamma}
\gamma = \frac{m g_{\rm 1D}}{\hbar^2 n_{\rm 1D}} 
.
\end{eqnarray}
We explain now why this model in the limit $|\gamma|\rightarrow\infty$ yields exactly Girardeau's solution \cite{girardeau:60}, irrespective of whether the interactions are elastic or inelastic.

The delta potential in Eq.\ (\ref{H-LL}) can be replaced by the following boundary condition at positions where the relative coordinate $x_{ij}=x_i-x_j$ vanishes \cite{lieb:63}
\begin{eqnarray}
\label{boundary}
\left.\frac{d\psi}{dx_{ij}}\right|_{x_{ij} \rightarrow0^+} - \left.\frac{d\psi}{dx_{ij}}\right|_{x_{ij}\rightarrow0^-}
= \frac{m g_{\rm 1D}}{\hbar^2} \left.\psi\right|_{x_{ij}=0}
.
\end{eqnarray}
Bosonic symmetry implies that on the left-hand side, the first term equals minus the second term. Expanding $\psi$ in a power series for $x_{ij}\rightarrow0$, Eq.\ (\ref{boundary}) can be rewritten as
\begin{eqnarray}
\label{boundary-ver2}
\psi(x_1,...,x_N) \propto \frac{2\hbar^2}{m g_{\rm 1D}} \linebreak[1] + |x_{ij}| +{\cal O}(x_{ij}^2)
,
\end{eqnarray}
where the proportionality contains the dependence on all remaining coordinates.

The energy eigenstates of this model can be divided into two classes: gaseous states and bound states. As an example of a bound state, we consider a state where ${\rm Re}(\gamma)<0$ and where particles $i$ and $j$ are bound. Here
\begin{eqnarray}
\psi \propto \exp\left( \frac{m g_{\rm 1D}}{2\hbar^2} |x_{ij}| \right)
\end{eqnarray}
for $x_{ij}\rightarrow 0$, which obviously meets the condition (\ref{boundary-ver2}). An expansion in a basis of plane waves $e^{ikx_{ij}}$ shows that this state contains imaginary momentum components that diverge for $\gamma\rightarrow-\infty$. Bound states exist only if ${\rm Re}(\gamma)<0$. The case of real and negative $\gamma$ was studied, {\it  e.g.}, in Refs.\ \cite{lieb:63,mcguire:64,calogero:75,castin:01,astrakharchik:05,calabrese:07}.

For real $\gamma$, two characteristic properties can be used to distinguish between bound and gaseous states: First, the momenta of all particles are real valued for gaseous states, whereas at least one momentum is complex for bound states. Second, for $\gamma\rightarrow\pm\infty$ all momenta converge to finite values for gaseous states, whereas at least one momentum diverges to $\pm i \infty$ for bound states. 

We consider the case of complex $\gamma$, where all momenta are usually complex so that the first characteristic property becomes useless, but the second one remains useful, namely the convergence or divergence of the momenta for $|\gamma|\rightarrow\infty$ can still be used to distinguish between gaseous and bound states. This distinction implies that $\psi$ and $d\psi/dx_{ij}$ remain finite for $|\gamma|\rightarrow \infty$ for all gaseous states. Hence, the left-hand side of Eq.\ (\ref{boundary}) remains finite for $|\gamma|\rightarrow \infty$, so that the right-hand side must remain finite, too. With $g_{\rm 1D}$ diverging, the right-hand side implies that
\begin{eqnarray}
\label{psi-vanishes}
\psi|_{x_{ij}=0}\rightarrow 0 \quad \mbox{ for } \quad |\gamma|\rightarrow \infty
\end{eqnarray}
for all gaseous states. The same result can be obtained from Eq.\ (\ref{boundary-ver2}), because the term proportional to $|x_{ij}|$ must remain finite so that the proportionality factor must remain finite. With $g_{\rm 1D}$ diverging, the constant term vanishes.

Yet another way to obtain Eq.\ (\ref{psi-vanishes}) is to consider a right eigenvector of the effective Hamiltonian $H_{\rm eff}|\psi_m \rangle= E_m |\psi_m \rangle$. If $|g_{\rm 1D} \psi(x_{ij}=0)| \rightarrow \infty$ for at least one relative coordinate $x_{ij}$, then Eq.\ (\ref{H-LL}) shows that ${\rm Im}(E)\rightarrow-\infty$, so that the state decays with infinite speed. We decompose $|\psi \rangle = \sum_m c_m |\psi_m \rangle$ and see that after any nonzero time only those eigenstates with finite $g_{\rm 1D} \psi(x_{ij}=0)$ survive. Again, this yields Eq.\ (\ref{psi-vanishes}). The other states are bound states. They vanish immediately and no longer contribute to the state or to the dynamics of the system.

In the limit $|\gamma|\rightarrow\infty$, the interaction for all gaseous states is fully described by the boundary condition (\ref{psi-vanishes}). The crucial point is that this boundary condition is independent of whether $\gamma$ is real or complex, and it is precisely this boundary condition that yields a Tonks-Girardeau gas. Hence, the wave functions of all gaseous states turn exactly into Girardeau's solutions \cite{girardeau:60} in the limit $|\gamma|\rightarrow\infty$, irrespective of whether $\gamma$ is real or complex. Attraction, repulsion, and dissipation all produce the Tonks-Girardeau gas in the limit of infinite interaction strength.

The above discussion shows that an initially uncorrelated state subject to strong dissipation will experience a rapid initial decay. Some fraction of the population will survive and this remaining population will be in the Tonks-Girardeau subspace, where the loss is slow. The loss will prevent the system from subsequently moving far away from the Tonks-Girardeau subspace. This freezing of the population can be understood intuitively with an analogy from classical optics \cite{duerr:0810.2217} or as a manifestation of the quantum Zeno effect \cite{syassen:08,garcia:09}. The focus of our study is the behavior of the system once it has reached the Tonks-Girardeau subspace, not the rapid initial decay. The experiment in Ref.\ \cite{syassen:08} started from a strongly correlated state already, so that no rapid initial decay occurred.

\subsection{Lieb-Liniger Solution}
Here, we summarize the central results of Lieb and Liniger \cite{lieb:63} insofar as they are required for the following discussion. Derivations for all results of this section can be found in Ref.\ \cite{lieb:63}. Following these derivations, one can show that these results also apply to the case of complex $g_{\rm 1D}$. The Lieb-Liniger solution is based on a Bethe ansatz \cite{lieb:63}
\begin{eqnarray}
\psi(x_1,...,x_N) &=& \sum_P a(P) \exp \left(i \sum_{j=1}^N x_j k_{P(j)}\right)
,
\end{eqnarray}
which holds only for $0\leq x_1 \leq x_2 \leq ... \leq x_N \leq L$. For other values of the coordinates, the solution is obtained from bosonic symmetry and periodic boundary conditions. The sum extends over all permutations $P$ of the numbers $1,...,N$. The parameters $k_1,...,k_N$ need to be determined. Once they are known, the amplitudes $a(P)$ can be calculated easily \cite{lieb:63}. The Bethe ansatz yields a nonzero solution for $\psi$ only if all $k_j$ are mutually different \cite{lieb:63}. The energy eigenvalue of this solution is
\begin{eqnarray}
\label{LL-energy}
E = \frac{\hbar^2}{2m} \sum_{j=1}^N k_j^2
.
\end{eqnarray}

Lieb and Liniger characterize the solutions by quantum numbers $n_1,...,n_{N-1}$, which are positive integers. For a given set of these quantum numbers the following set of coupled implicit equations is to be solved for $j=1,...,N-1$ \cite{lieb:63}
\begin{eqnarray}
\label{n-j}
(k_{j+1}-k_j) L = 2\pi n_j + \sum_{s=1}^N (\theta_{s,j} - \theta_{s,j+1})
,
\end{eqnarray}
where $\theta_{i,j}$ abbreviates \cite{lieb:63}
\begin{eqnarray}
\label{theta-ij-arctan}
\theta_{i,j} = -2 \arctan \left( \frac{k_i-k_j}{\gamma n_{\rm 1D}} \right)
\end{eqnarray}
with $|{\rm Re}(\theta_{i,j})| \leq \pi$. This set of $N-1$ implicit equations determines only the $N-1$ quantities $k_{j+1}-k_j$. There is one remaining degree of freedom. It can be expressed in terms of the quantum number \cite{lieb:63}
\begin{eqnarray}
\label{n-0}
n_0= \frac{L}{2\pi} \sum_{j=1}^N k_j
\end{eqnarray}
which must also be an integer. For a given set of quantum numbers $n_1,...,n_{N-1}$, the value of $n_0$ is constrained by the additional condition \cite{lieb:63}
\begin{eqnarray}
\label{periodic-boundary}
(-1)^{N-1} e^{-i k_1 L} = \exp\left( i \sum_{s=1}^N \theta_{s,1} \right)
.
\end{eqnarray}
If one restricts $n_0$ to the range \cite{lieb:63}
\begin{eqnarray}
\label{n0-range}
- \frac N2 < n_0 \leq \frac N2
,
\end{eqnarray}
then there is exactly one value $n_0$ for each combination of the $n_1,...,n_{N-1}$.

Adding $N$ to $n_0$ adds a momentum of $2\pi/L$ to each particle in the system. This corresponds to a center-of-mass motion of the gas as a whole, which is trivial and of little interest. The values of $n_0$ in the range (\ref{n0-range}), however, correspond to different internal quantum numbers and thus need to be taken into account \cite{lieb:63}.

While the Bethe ansatz does yield solutions for complex $\gamma$, it remains unclear whether a complete set of solutions is obtained. Our numerical results, discussed further below, suggest that each solution for real (positive or negative) $\gamma$ continuously deforms into exactly one solution for complex $\gamma$, thus suggesting that completeness for real $\gamma$ and complex $\gamma$ are closely linked. But a solid proof of completeness for complex $\gamma$ is beyond the scope of the present paper. On the other hand, the completeness issue has only limited relevance for the experiment discussed in Ref.\ \cite{syassen:08}, because here only the time evolution of specifically-prepared gaseous initial states is to be tracked.

\subsection{Fermionized States in the Limit of Strong Interactions}
We now consider the limit $|\gamma| \rightarrow \infty$. We expand the $\arctan$ in Eq.\ (\ref{theta-ij-arctan}) in a power series and keep only the lowest order. This requires that $(k_i-k_j)/n_{\rm 1D}\gamma$ vanishes for large $|\gamma|$. A sufficient condition is that all $k_j$ remain finite, which defines the gaseous states.

For any large but finite $|\gamma|$, the gaseous states can be divided into two classes: first, the so-called fermionized states for which $|(k_i-k_j)/n_{\rm 1D}\gamma|\ll 1$ so that we can expand the $\arctan$ and, second, the other states which must have very large $k_j$ so that they are highly excited. In the limit $|\gamma| \rightarrow \infty$ all gaseous states become fermionized.

Insertion of this expansion into Eq.\ (\ref{n-j}) yields
\begin{eqnarray}
\label{kj+1-kj}
k_{j+1}-k_j = \frac{2\pi}L \; \frac{\gamma}{\gamma+2} n_j + {\cal O}(|\gamma|^{-3})
.
\end{eqnarray}
Iterating this equation, we obtain for $j=2,...,N$
\begin{eqnarray}
\label{k-j}
k_j = k_1 + \frac{2\pi}L \frac{\gamma}{\gamma+2} \sum_{s=1}^{j-1} n_s + {\cal O}(|\gamma|^{-3})
.
\end{eqnarray}
Insertion into Eq.\ (\ref{n-0}) yields
\begin{eqnarray}
\label{k-1}
k_1 = \frac{2\pi}{N L} \left( n_0 - \frac{\gamma}{\gamma+2} \sum_{s=1}^{N-1} (N-s) n_s \right)
+ {\cal O}(|\gamma|^{-3})
.
\end{eqnarray}
Insertion into Eq.\ (\ref{periodic-boundary}) yields that
\begin{eqnarray}
\label{get-n-0}
\frac 1N \left( n_0 - \sum_{s=1}^{N-1} (N-s) n_s \right) + \frac{N-1}{2} +{ \cal O}(|\gamma|^{-1})
\end{eqnarray}
must be an integer, thus fixing $n_0$ modulo $N$. As $n_0$ must generally be an integer, the term ${\cal O}(|\gamma|^{-1})$ must vanish automatically. For given values of $\gamma$ and $n_1,...,n_{N-1}$, we thus obtain $n_0$ and all the $k_j$.

\subsection{Ground State in the Limit of Strong Interactions}
These general results for fermionized states can be used to calculate the ground state properties. The ground state of the gas phase is characterized by \cite{lieb:63}
\begin{eqnarray}
\label{all-n=1}
n_1=n_2=...=n_{N-1}=1
.
\end{eqnarray}
Eqs.\ (\ref{k-j}) - (\ref{get-n-0}) yield $n_0=0$ and
\begin{eqnarray}
k_j = \frac{2\pi}L \frac{\gamma}{\gamma+2} \left( j -\frac{N+1}2 \right)
+ {\cal O}(|\gamma|^{-3})
.
\end{eqnarray}
According to Eq.\ (\ref{LL-energy}) the ground state energy is
\begin{eqnarray}
\label{E-ground-state}
E = N \frac{\pi^2 \hbar^2 n_{\rm 1D}^2}{6 m} \left( \frac{\gamma}{\gamma+2} \right)^2 \left(1-\frac1{N^2} \right)
+ {\cal O}(|\gamma|^{-3})
.
\end{eqnarray}
This result was previously derived by Lieb and Liniger \cite{lieb:63} in an alternative way for the case of real $\gamma$ and in the limit $N\rightarrow \infty$. Our above calculation generalizes this result to complex $\gamma$ and finite $N$.

Expansion to lowest order in $1/\gamma$, taking the limit $N\rightarrow\infty$, and insertion into Eq.\ (\ref{dN-dt-Im-E}) yield the particle loss rate at short times
\begin{eqnarray}
\label{dNdt-ground-state}
\left. \frac{d N}{dt} \right|_{t=0} = - N \frac{8\pi^2 \hbar n_{\rm 1D}^2}{3m} \; {\rm Im}\left(\frac{1}{\gamma}\right)
+{\cal O}(|\gamma|^{-2})
.
\end{eqnarray}
We use ${\rm Im}(1/\gamma)=-{\rm Im}(\gamma)/|\gamma|^2$ and compare with Eqs.\ (\ref{dndt-1D}) and (\ref{K-1D}). This yields one of the central results of this paper
\begin{eqnarray}
\label{g2-ground-state}
g^{(2)} = \frac{4\pi^2}{3|\gamma|^2}
.
\end{eqnarray}
For the special case of real $\gamma$, this result was previously obtained in Ref.\ \cite{gangardt:03}. Note that $g^{(2)} \rightarrow 0$ for $|\gamma|\rightarrow \infty$, irrespective of whether $\gamma$ is real or complex. This is a consequence of the boundary condition (\ref{psi-vanishes}).

\begin{figure}[t!]
\includegraphics[width=0.9\columnwidth]{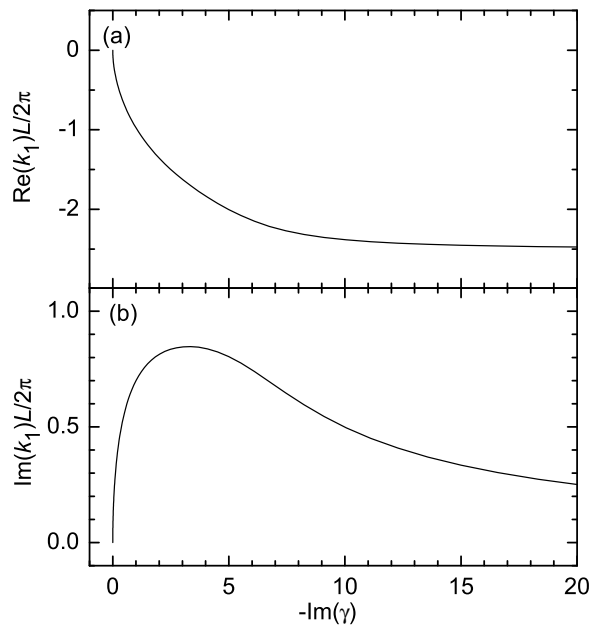}
\caption{
\label{fig-k-vs-gamma}
Real and imaginary parts of $k_1$ for the $N=6$ ground state with ${\rm Re}(\gamma)=0$. The same data are plotted in a different way as the top-left curve in Fig.\ \ref{fig-complex-k-plane}(a).
}
\end{figure}

Eq.\ (\ref{dNdt-ground-state}) reveals that the loss rate depends on the scattering properties only in terms of ${\rm Im}(1/\gamma) \propto {\rm Im}(1/g_{\rm 1D})$. We can rewrite Eq.\ (\ref{g-1D}) as
\begin{eqnarray}
\frac1{g_{\rm 1D}} = \frac{m a_\perp^2}{2\hbar^2} \left( \frac1a + \frac{\zeta\left(1/2\right)}{\sqrt2 \; a_\perp} \right)
\end{eqnarray}
and find that ${\rm Im}(1/g_{\rm 1D}) \propto {\rm Im}(1/a)$, irrespective of whether the condition $|a|\ll a_\perp$ is met or not. A possible alteration of the scattering process in the regime where the condition $|a|\ll a_\perp$ is not met affects $g_{\rm 1D}$ in such a way that the loss rate in the fermionized regime is unchanged. In the fermionized regime, the dependence of the loss rate on the scattering length is always given by
\begin{eqnarray}
\label{dNdt-Ima}
\left. \frac{d N}{dt} \right|_{t=0} \propto {\rm Im}\left(\frac{1}{a}\right)
\end{eqnarray}
Note that in a case where an additional lattice is applied along the one dimension, the loss rate also obeys Eq.\ (\ref{dNdt-Ima}), see appendix \ref{app-experiment} and Refs.\ \cite{syassen:08,garcia:09}.

\subsection{Numerical Solutions}

Solutions of the Lieb-Liniger model outside the limit $|\gamma|\rightarrow\infty$ can be found numerically. To this end, we numerically solve the set of coupled Eqs.\ (\ref{n-j}) and (\ref{theta-ij-arctan}) with the standard root-finding algorithm in {\em Mathematica}. Computation time issues restrict this approach to reasonably small $N$. It is not necessary to choose a specific value of $L$ for the numerical calculations, because the relevant equations can be written in terms of the dimensionless parameters $N$, $\gamma$, and $kL$.

\begin{figure}[t!]
\includegraphics[width=0.9\columnwidth]{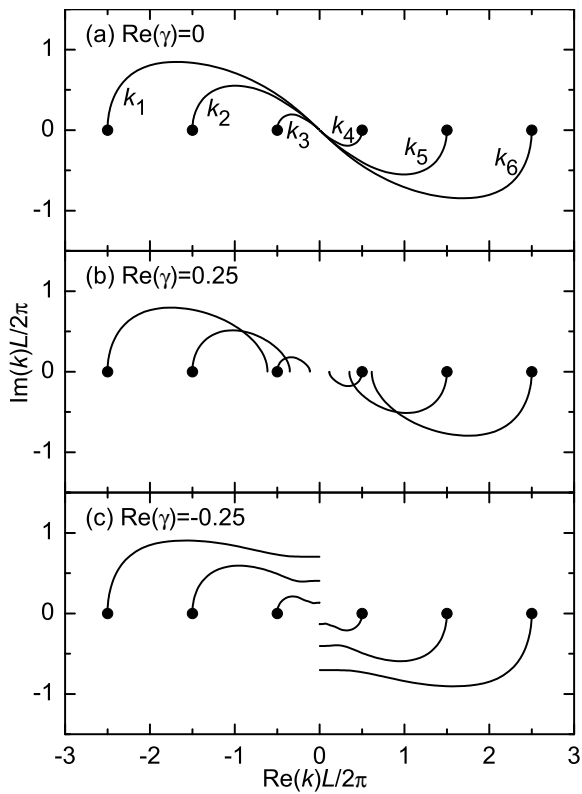}
\caption{
\label{fig-complex-k-plane}
Parametric plot, showing the $k_j$ in the complex $k$ plane for the $N=6$ ground state. Each curve shows a trajectory traced by $k_j$ when scanning the parameter ${\rm Im}(\gamma)$ from 0 to $-\infty$. In all cases, the curves begin for ${\rm Im}(\gamma)=0$ somewhere not too far from the coordinate origin and converge for ${\rm Im}(\gamma)\rightarrow-\infty$ to Girardeau's solution (shown as filled circles). (a) ${\rm Re}(\gamma)=0$. For $\gamma=0$ all $k_j=0$. (b) ${\rm Re}(\gamma)=0.25$. For ${\rm Im}(\gamma)=0$, the $k_j$ are approximately equally spaced on the real axis. (c) ${\rm Re}(\gamma)=-0.25$. For ${\rm Im}(\gamma)=0$, the $k_j$ are approximately equally spaced on the imaginary axis, indicating a bound state with an energy that is real and negative. However, as $|{\rm Im}(\gamma)|$ grows and becomes larger than $|{\rm Re}(\gamma)|$, the finite ${\rm Re}(\gamma)$ becomes irrelevant and Girardeau's solution is reached for $|\gamma|\rightarrow\infty$.
}
\end{figure}

We illustrate these solutions for the ground state characterized by Eq.\ (\ref{all-n=1}) with $N=6$ particles. Figure \ref{fig-k-vs-gamma} shows the real and imaginary parts of one of the parameters, namely $k_1$, as a function of $-{\rm Im}(\gamma)$. An alternative way of displaying the same data is to show the trajectory that $k_1$ follows in the complex $k$ plane when ${\rm Im}(\gamma)$ is scanned from 0 to $-\infty$. Such plots are shown in Fig.\ \ref{fig-complex-k-plane} for all $k_j$. Parts (a), (b), and (c) each correspond to one fixed value of ${\rm Re}(\gamma)$. For ${\rm Im}(\gamma)\rightarrow - \infty$ the solutions in all parts of the figure converge to Girardeau's solution (shown as filled circles). The important result is that ${\rm Re}(\gamma)$ becomes irrelevant for $|{\rm Re}(\gamma)| \ll |{\rm Im}(\gamma)|$ and that Girardeau's solution is reached in any case.

\begin{figure}[t!]
\includegraphics[width=0.9\columnwidth]{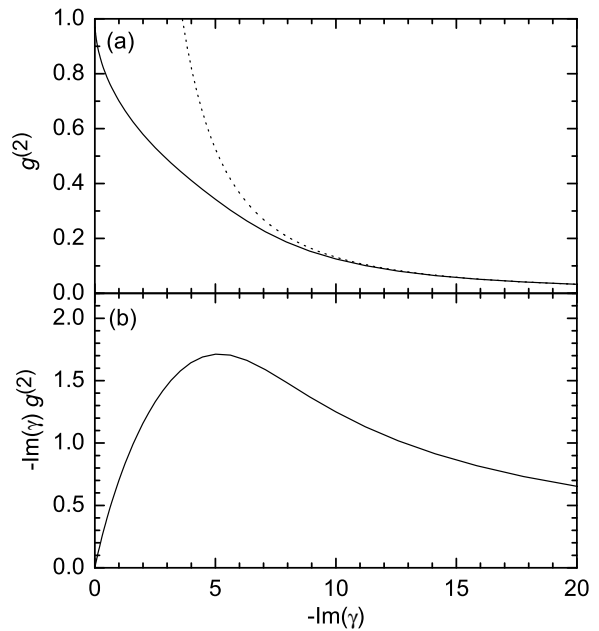}
\caption{
\label{fig-g2}
Pair correlation function and loss rate of the ground state as a function of the dissipation strength for ${\rm Re}(\gamma)=0$ and $N=100$. (a) The pair correlation function $g^{(2)}$ (solid line) equals $(N-1)/N\approx 1$ at $\gamma=0$ and vanishes for $\gamma\rightarrow -i\infty$. For $|\gamma|\gg 1$ it is well approximated by Eq.\ (\ref{g2-ground-state}) (dotted line). (b) The dimensionless quantity $-{\rm Im} (\gamma) g^{(2)}$ is proportional to the particle loss rate, see Eq.\ (\ref{dndt-1D-gamma}). For small $|\gamma|$ the loss rate increases linearly with ${\rm Im}(\gamma)$ because $g^{(2)}$ is approximately constant. But for large $|\gamma|$, the suppression of $g^{(2)}\propto |\gamma|^{-2}$ dominates and the loss rate vanishes like $1/|\gamma|$. This is remarkable: the loss suppresses itself.
}
\end{figure}

The values of $k_j$ obtained in such numerical calculations can be used to extract $g^{(2)}$ in the same way as in the derivation of Eq.\ (\ref{g2-ground-state}). Results for the ground state are shown in Fig.\ \ref{fig-g2}(a). $g^{(2)}$ depends on $N$, but converges in the thermodynamic limit. The solid line in Fig.\ \ref{fig-g2}(a) shows the case $N=100$. On the scale shown here it cannot be distinguished from the case $N=10$ (not shown). This suggests that Fig.\ \ref{fig-g2}(a) shows a reasonable approximation to the thermodynamic limit. A thorough analysis of finite-size effects is beyond the scope of this work.

The loss rate Eq.\ (\ref{dndt-1D}) can be rewritten as
\begin{eqnarray}
\label{dndt-1D-gamma}
\frac{d n_{\rm 1D}}{dt} = \frac{2\hbar n_{\rm 1D}^3}{m} {\rm Im}(\gamma) g^{(2)}
.
\end{eqnarray}
The dependence on $\gamma$ is given by the dimensionless expression ${\rm Im}(\gamma) g^{(2)}$. This quantity is displayed in Fig.\ \ref{fig-g2}(b), showing that for $-{\rm Im}(\gamma)\gg 1$ the loss suppresses itself. The experiment in Ref.\ \cite{syassen:08} was performed for $|\gamma|=11$ which is in this regime.

\section{Loss at Longer Times}

So far we studied only the loss rate at short times. A detailed study of loss rates at longer times would require a numerical solution of the master equation including the term ${\cal J}(\rho)$. This is beyond the scope of the present paper, but for the case with a lattice applied along the one dimension, results of such a numerical calculation are reported in Ref.\ \cite{garcia:09}.

Instead of solving the master equation, we now present a fairly simple estimate for the temporal evolution of the loss rate. During the loss, a spread in particle number builds up, but this spread is expected to reach a Poisson distribution, which is narrow enough that it suffices to consider only the mean value of the particle number. As the loss proceeds, the particle number changes and so does the ground-state energy $E_g$ in Eq.\ (\ref{E-ground-state}). In the Lieb-Liniger model with periodic boundary conditions, $E_g \propto N^3$. In a more realistic model with a harmonic trap along the one dimension, the scaling is $E_g \propto N^2$, which we use in the following. A simple estimate can be obtained when assuming that the average kinetic ({\it i.e.}, real) part of the energy per particle is preserved by the loss, so that ${\rm Re}(E)/N$ remains constant. We assume that $E=E_g$ at $t=0$. As $N$ decreases, ${\rm Re}(E)/{\rm Re}(E_g)$ will then increase according to ${\rm Re}[E(t)]/{\rm Re}[E_g(t)]=N(0)/N(t)$. The experiment in Ref.\ \cite{syassen:08} monitored a decrease of $N$ by a factor of $\sim 2$. Our simple estimate predicts an increase of ${\rm Re}[E(t)]/{\rm Re}[E_g(t)]$ by the same factor of 2. We conclude that while the loss occurs, the energy of the system evolves somewhat away from the energy of the ground state, but not terribly far.

This simple estimate is based on the {\em ad hoc} assumption that ${\rm Re}(E)/N$ is preserved by the loss. We now study how realistic this assumption is. As the loss takes the system into the regime ${\rm Re}[E(t)]>{\rm Re}[E_g(t)]$, we need to study the excited states of the gas phase. We restrict our considerations to the fermionized regime. In a first step, we ignore the condition (\ref{periodic-boundary}) and set $n_0=0$. Eqs.\ (\ref{LL-energy}), (\ref{k-j}), and (\ref{k-1}) yield
\begin{eqnarray}
E|_{n_0=0} &=& \frac{\hbar^2}{2m} \left( \frac{2\pi}{L} \frac{\gamma}{\gamma+2} \right)^2
f(n_1,...,n_{N-1})
,
\end{eqnarray}
where
\begin{eqnarray}
f =\sum_{j=1}^N \left( \sum_{s=1}^{j-1} n_s - \frac 1N \sum_{s=1}^{N-1} (N-s) n_s\right)^2
.
\end{eqnarray}
As the ground state truly has $n_0=0$ we obtain
\begin{eqnarray}
E|_{n_0=0} = \frac{f(n_1,...,n_{N-1})}{f(1,...,1)} \; E_g
,
\end{eqnarray}
where $E_g$ denotes the ground-state energy. Obviously $f$ is always real. Hence, 
\begin{eqnarray}
\label{E-n=0-E0}
\frac{{\rm Im}(E|_{n_0=0})}{{\rm Re}(E|_{n_0=0})} = \frac{{\rm Im}(E_g)}{{\rm Re}(E_g)}
.
\end{eqnarray}
We will show now that the correction for $n_0\neq0$ is negligible. To this end, we note that according to Eqs.\ (\ref{k-j}) and (\ref{k-1}), $n_0$ shifts all $k_j$ by $2\pi n_0/NL$. As a result, the energy becomes
\begin{eqnarray}
E = E|_{n_0=0} + \frac{\hbar^2}{2m N} \left( \frac{2\pi n_0}{L}\right)^2
.
\end{eqnarray}
Eq.\ (\ref{n0-range}) yields $|n_0|/L\leq n_{\rm 1D}/2$. In the thermodynamic limit ($N\rightarrow \infty$ at fixed $n_{\rm 1D}$), the correction is of order $E - E|_{n_0=0} \leq n_{\rm 1D}^2 {\cal O}(1/N)$. On the other hand, $E|_{n_0=0}\geq E_g \propto N n_{\rm 1D}^2$. Hence,
\begin{eqnarray}
\frac{E}{E|_{n_0=0}} = 1 + {\cal O}(N^{-2})
.
\end{eqnarray}
Therefore the correction due to $n_0$ is negligible and Eq.\ (\ref{E-n=0-E0}) becomes
\begin{eqnarray}
\frac{{\rm Im}(E)}{{\rm Re}(E)} = \frac{{\rm Im}(E_g)}{{\rm Re}(E_g)}
.
\end{eqnarray}
This shows that the ratio of the real to the imaginary part of $E$ is identical for all fermionized states, independent of $n_1,...,n_{N-1}$. Combined with Eq.\ (\ref{dN-dt-Im-E}) we obtain that the loss rate of a state is proportional to the kinetic ({\it i.e.}, real) part of the energy of a state.

This has two important consequences for the understanding of the loss at longer times. First, it shows that an increase of ${\rm Re}[E(t)]/{\rm Re}[E_g(t)]$ by a given factor leads to an increase of the loss rate compared to that of the ground state by the same factor. Second, it shows that the loss preferentially depletes states with large energy. Hence, the above estimate assuming that the loss leaves the average kinetic energy per particle unchanged is too pessimistic. In reality, the ratio ${\rm Re}[E(t)]/{\rm Re}[E_g(t)]$ increases even more slowly than predicted by the simple estimate above. The kinetic energy and loss rate of the system will thus remain fairly close to those of the ground state if the loss is allowed to evolve only for a time where the particle number decays by a factor of $\sim 2$ as in Ref.\ \cite{syassen:08}.

\section{Conclusion}
In conclusion, we showed that strong inelastic interactions produce a Tonks-Girardeau gas, much as strong elastic interactions. We derived an analytic expression for the pair correlation function of the fermionized ground state. Numerical results illustrate the behavior of the ground state for small and medium dissipation strength. These calculations also show that the loss suppresses itself. Finally, we presented a simple estimate for the evolution of the loss rate for longer times.

\appendix

\section{Derivation of the Master Equation}
\label{app-master}
\subsection{General Considerations}
In this appendix we derive the master equation (\ref{master}) with the dissipator (\ref{D}) for the description of loss of particles from a many-body system due to ultracold inelastic two-body collisions. The system of interest $S$ consists of trapped motional states in the initial internal state. Inelastic collisions couple this system $S$ to other states, which we call the reservoir $R$. The reservoir states can represent all possible internal states. The motional parts of the reservoir states are untrapped because we assumed that all particles involved in an inelastic collision quickly escape from the trap. Hence the energy spectrum of these untrapped states is a continuum.

The total system consisting of $S$ and $R$ is represented by a density matrix $\chi$. As we are not interested in the reservoir, we can trace out the reservoir degrees of freedom and obtain a density matrix $\rho$ that describes only the system of interest $S$. We assume that the reservoir is in the vacuum state initially and we use a Born-Markov approximation \cite{carmichael:99} for the reservoir modes, {\it i.e.}, we assume that the reservoir has no memory of the past. This is justified because the reservoir consists of a continuum of modes that quickly dephase. As a result, the reservoir returns to the vacuum state on a time scale, which is much faster than all time scales in the system of interest $S$. Using very general arguments one can show \cite{carmichael:99,breuer:02} that such a Born-Markov approximation yields a Markovian quantum master equation
\begin{eqnarray}
\label{master-app}
\hbar \frac{d\rho}{dt} &=& - i [H_1,\rho] + {\cal D}(\rho)
\end{eqnarray}
for the evolution of the density matrix $\rho$ describing only the system of interest $S$. This master equation contains a Hamiltonian $H_1$ and a so-called dissipator $\cal D$, which models the irreversible part of the time evolution and can always be written in a Lindblad form \cite{carmichael:99,breuer:02}
\begin{eqnarray}
{\cal D}(\rho) = \frac{\hbar}{2} \sum_k \gamma_k \left( 2 A_k \rho A_k^\dag - A_k^\dag A_k \rho - \rho A_k^\dag A_k \right)
,
\end{eqnarray}
where the $\gamma_k\geq0$ are relaxation rate coefficients and the $A_k$ are called Lindblad operators.

The rest of this appendix is devoted to deriving expressions for $H_1$ and $\cal D$ for our particular system. Given the form of the elastic collision term in Eq.\ (\ref{H-0}), the final result for the Lindblad operator $A_k=\Psi^2({\bf x})$ in Eq.\ (\ref{D}) is not terribly surprising. But the result for the rate coefficient $\gamma_k=-{\rm Im}(g_{\rm 3D})/\hbar$ is certainly less obvious, in particular as far as possible factors of 2 are concerned. We now present a rigorous derivation for this dissipator, including the rate coefficient. The calculation proceeds in three steps. First, we determine a contact potential that can be used to describe inelastic scattering, second, we derive the corresponding Hamiltonian in second quantization and, third, we eliminate the reservoir degrees of freedom.

\subsection{Contact Potential for Inelastic Scattering}
For the first step, we consider a scattering process with two distinguishable incoming particles and assume, for simplicity, that after the collision, there are two outgoing particles with the same mass as the incoming particles. We further assume that the incoming wave is an $s$ wave and that the potential is spherically symmetric, so that the outgoing wave is also an $s$ wave. The scattering state $|\psi\rangle$ at large relative distance of the particles is
\begin{eqnarray}
\label{r-psi}
\langle {\bf r}|\psi\rangle
&=& \frac{e^{-ik_{\alpha\beta} r}}r |\alpha\beta\rangle 
\nonumber \\ &&
- \sum_{\alpha'\beta'} \sqrt{\frac{k_{\alpha\beta}}{k_{\alpha'\beta'}}} S_{\alpha'\beta',\alpha\beta} \frac{e^{ik_{\alpha'\beta'} r}}r |\alpha'\beta'\rangle
,
\end{eqnarray}
where ${\bf r}$ is the relative coordinate, $S_{\alpha'\beta',\alpha\beta}$ is a matrix element of the scattering matrix, and the internal states of the particles before and after the collision are $|\alpha\beta\rangle=|\alpha\rangle\otimes|\beta\rangle$ and $|\alpha'\beta'\rangle$ with corresponding wave vectors for the relative motion ${\bf k}_{\alpha\beta}$ and ${\bf k}_{\alpha'\beta'}$, respectively. The wave function (\ref{r-psi}) can be regarded as a generalization of Eq.\ (5) in Ref.\ \cite{duerr:05} to inelastic $s$-wave scattering. The fact that the outgoing particle flux is proportional to $k_{\alpha'\beta'}$ makes it necessary to include the factor $\sqrt{k_{\alpha\beta} / k_{\alpha'\beta'}}$ in Eq.\ (\ref{r-psi}), in order to maintain the fact that the $S$-matrix is unitary if the number of particles is conserved in the collision.

We now replace the interparticle interaction potential by a contact potential. The wave function (\ref{r-psi}) that is usually valid for $r\to \infty$ only, then becomes valid for all $r$. We choose a contact potential $V^{(2)}=g \delta_{\rm reg}^{(3)}$ with a part $g$ that acts only on the internal state and a part $\delta_{\rm reg}^{(3)}$ that acts only on the spatial part of the wave function. For the latter, we use the regularized delta function $\langle {\bf r}|\delta_{\rm reg}^{(3)}|\psi\rangle=\delta^{(3)}({\bf r})\frac{\partial}{\partial r}(r\psi)$ (see, {\it e.g.}, Refs.\ \cite{huang:87Sec10.5,busch:98}). In addition, we assume that $g$ is Hermitian.

We expand the wave function (\ref{r-psi}) in a power series for $r\to0$ and spatially integrate the Schr\"odinger equation $\langle{\bf r}|(H-E)|\psi\rangle=0$ over a sphere centered around $r=0$ with radius $\epsilon\rightarrow 0^+$. We use $\nabla^2\frac1r=-4\pi \delta^{(3)}({\bf r})$ and $\int_0^\epsilon dr 4\pi r^2 \delta^{(3)}({\bf r}) =1$. We neglect terms ${\cal O}(\epsilon)$. This yields for all $\alpha\beta\alpha'\beta'$
\begin{eqnarray}
\label{coeff-zero}
&&
\sum_{\alpha''\beta''} g_{\alpha'\beta',\alpha''\beta''} \sqrt{k_{\alpha''\beta''}} (S_{\alpha''\beta'',\alpha\beta}+\delta_{\alpha''\alpha}\delta_{\beta''\beta}) 
=  \nonumber \\
&& \quad = i \frac{2\pi\hbar^2}{\mu \sqrt{k_{\alpha'\beta'}}} (S_{\alpha'\beta',\alpha\beta} -\delta_{\alpha'\alpha}\delta_{\beta'\beta}) 
\end{eqnarray}
with the reduced mass $\mu=m/2$ and the Kronecker symbol $\delta$. We do not attempt to solve this linear system for the variables $g_{\alpha'\beta',\alpha\beta}=\langle \alpha'\beta'|g|\alpha\beta\rangle$ in full generality. Instead, we decide to fulfill Eq.\ (\ref{coeff-zero}) for only one specific initial state $|\alpha\beta\rangle=|ii\rangle$. We thus assume
\begin{eqnarray}
\label{g-zero}
g_{\alpha'\beta',\alpha\beta} = 0 \mbox{ if } (\alpha',\beta')\neq (i,i) \mbox{ and } (\alpha,\beta) \neq (i,i)
.
\end{eqnarray}
This yields
\begin{eqnarray}
\label{g-alpha-beta-ii}
g_{\alpha\beta,ii} &=& i \frac{2\pi \hbar^2 S_{\alpha\beta,ii}}
{\mu (S_{ii,ii}+1) \sqrt{k_{\alpha\beta}k_{ii}} } \mbox{ if } (\alpha,\beta)\neq (i,i)
\quad
\\
\label{g-Hermitian}
g_{ii,\alpha\beta} &=& g_{\alpha\beta,ii}^*
\\
g_{ii,ii} &=& 
- \frac{4\pi\hbar^2 {\rm Im}(S_{ii,ii})}{\mu k_{ii} |S_{ii,ii}+1|^2} 
,
\end{eqnarray}
where we used that unitarity of the $S$-matrix implies $\sum_{\alpha\beta}|S_{\alpha\beta,ii}|^2=1$.

The scattering phase $\eta$, defined by $S_{ii,ii}=e^{2i\eta}$, can be used to rewrite the above results in the compact form
\begin{eqnarray}
\label{g-ii}
g_{ii,ii}
&=& - \frac{2\pi\hbar^2}{\mu} \ \frac{{\rm Re}(\tan\eta)}{k_{ii}}
\\
\label{sum-g-fi}
\sum_{\alpha\beta\neq ii} k_{\alpha\beta} |g_{\alpha\beta,ii}|^2 
&=& \left(\frac{2\pi \hbar^2}{\mu}\right)^2 \frac{{\rm Im}(\tan\eta)}{k_{ii}}
.
\quad
\end{eqnarray}
For ultracold collisions, the ratio $\tan\eta/k_{ii}$ becomes constant and is called the scattering length
\begin{eqnarray}
\label{a}
a= - \lim_{k_{ii}\to0} \frac{\tan\eta}{k_{ii}}
.
\end{eqnarray}

The above treatment applies to scattering of distinguishable particles, while what we are really interested in is scattering of identical bosons. Due to the linearity of the Schr\"odinger equation, scattering of identical bosons is closely related the scattering of distinguishable particles from the same potential. Our following approach is based on second quantization, which has bosonic symmetry built into it automatically, so that no separate treatment of scattering of identical bosons is needed here.

\subsection{Second Quantization}
The interactions described by the above contact potential can be translated to a Hamiltonian in second quantization for a many-body system. If a general two-body interaction potential $V^{(2)}({\bf x}_\mu,{\bf x}_\nu)$ is summed over all possible pairs of particles, one obtains $\sum_{\mu<\nu}V^{(2)}({\bf x}_\mu,{\bf x}_\nu)=
\frac12 \sum_{\mu\neq\nu}V^{(2)}({\bf x}_\mu,{\bf x}_\nu)$, where $\mu,\nu$ enumerate the particles. In second quantization, this yields a Hamiltonian \cite{schwabl:05}
\begin{eqnarray}
 H_V = \frac12 \sum_{jklm} \langle j,k| V^{(2)} |l,m \rangle  b_j^\dag  b_k^\dag  b_l  b_m
,
\end{eqnarray}
where the states $|j\rangle$ form a basis of single-particle states and $ b_j^\dag$ creates a boson in state $|j\rangle$.

We now apply this to our contact potential. As a basis of single particle states, we choose $|\alpha\rangle\otimes|{\bf k}\rangle $, where $|\alpha\rangle$ is an internal state and $|{\bf k}\rangle$ a momentum state with $\langle{\bf x}|{\bf k}\rangle=e^{i{\bf k}\cdot{\bf x}}L^{-3/2}$ in box quantization with quantization volume $L^3$. We denote the corresponding creation operator as $a^\dag_{\alpha,\bf k}$. A short calculation yields the scattering Hamiltonian for our contact potential
\begin{eqnarray}
\label{H-sc}
H_{\rm sc} 
&=& \sum_{\alpha'\beta'} \sum_{\alpha\beta} \sum_{{\bf kpq}} \frac{g_{\alpha'\beta',\alpha\beta}}{2L^3} 
a_{\alpha',{\bf k}+{\bf q}}^\dag a_{\beta',{\bf k}-{\bf q}}^\dag a_{\alpha,{\bf k}+{\bf p}} 
\nonumber \\ & \times &
a_{\beta,{\bf k}-{\bf p}} \underbrace{\int d^3r e^{-i({\bf k}+{\bf q})\cdot {\bf r}} \delta_{\rm reg}^{(3)}({\bf r}) e^{i({\bf k}+{\bf p})\cdot {\bf r}}}_{=1}
,
\end{eqnarray}
where $2\hbar{\bf k}$ is the center-of-mass momentum of a pair of bosons, and $\hbar{\bf p}$ and $\hbar{\bf q}$ are the relative momenta before and after the collision. We denote the corresponding kinetic energies as $\hbar \omega_k=(2\hbar k)^2/4m$, $\hbar \omega_p=\hbar^2 p^2/2\mu$, and $\hbar \omega_q=\hbar^2 q^2/2\mu$. Note that in Eq.\ (\ref{H-sc}) it makes no difference if we take $\delta_{\rm reg}^{(3)}({\bf r})$ or $\delta^{(3)}({\bf r})$.

Let us assume for a moment that the $g_{\alpha'\beta',\alpha\beta}$ are independent of ${\bf k}$, ${\bf p}$, and ${\bf q}$. $H_{\rm sc}$ can then be transformed into a simple expression in position representation
\begin{eqnarray}
H_{\rm sc} 
&=& 
\frac12 \sum_{\alpha'\beta'}\sum_{\alpha\beta} g_{\alpha'\beta',\alpha\beta} 
\nonumber \\ && \times
\int d^3x \Psi^\dag_{\alpha'} ({\bf x}) \Psi^\dag_{\beta'} ({\bf x}) \Psi_\alpha({\bf x})
\Psi_\beta({\bf x})
,
\end{eqnarray}
where the field operator $\Psi^\dag_{\alpha}({\bf x})=\sum_{{\bf k}} a_{\alpha,\bf k}^\dag \langle {\bf k}| {\bf x}\rangle$ creates a boson at position ${\bf x}$ in internal state $|\alpha\rangle$.

Note that the term in $H_{\rm sc}$ that describes elastic scattering in the initial internal state is proportional to $g_{ii,ii}$ from Eq.\ (\ref{g-ii}). Using Eq.\ (\ref{a}), we obtain exactly the elastic collision term in Eq.\ (\ref{H-0}). The resulting mean-field energy is small compared to the kinetic energies involved in the following calculation. Hence we can perform an approximation of independent rates of variation \cite{cohen-tannoudji:92}. This means that we include the elastic collision term in the final Eq.\ (\ref{H-0}), but we neglect its effect on $\cal D$ by setting $g_{ii,ii}=0$ in the following calculation.

\subsection{Eliminating the Reservoir}
Having determined the explicit form of $ H_{\rm sc}$, we now proceed to eliminate the reservoir modes. To this end, we split the total Hamiltonian of our problem into three parts: $H_S$ acting only on the system of interests $S$, $H_R$ acting only on the reservoir $R$, and $H_{SR}$ coupling $S$ and $R$. The single-particle Hamiltonian is thus decomposed into
\begin{eqnarray}
H_S+H_R
=
\sum_{\alpha \bf k} \left( \frac{\hbar^2{\bf k}^2}{2m} +\hbar \omega_\alpha \right) 
 a_{\alpha,\bf k}^\dag  a_{\alpha,\bf k}
,
\end{eqnarray}
where the term $\alpha=i$ belongs to $H_S$, all others to $H_R$. Here $\hbar \omega_\alpha$ is the internal energy in internal state $|\alpha\rangle$. We choose the zero of internal energies such that $\omega_i=0$.

For simplicity, we neglect collisions except if they make a transition between $S$ and $R$. Hence $H_{SR}= H_{\rm sc}$ with the $g_{\alpha'\beta',\alpha\beta}$ from Eqs.\ (\ref{g-zero})--(\ref{g-Hermitian}) and with the assumption $g_{ii,ii}=0$. We further simplify the model by excluding the possibility that a reservoir state contains any particles in the initial internal state $|i\rangle$. This is a simple way to reflect the fact that all particles involved in an inelastic collision are lost from the sample.

Our discussion throughout the rest of this section closely follows chapter 1 of Ref.\ \cite{carmichael:99}. The total system consisting of $S$ and $R$ is described by a density matrix $\chi(t)$ that obeys the von-Neumann equation $i\hbar\dot\chi=[H_S+H_R+H_{SR},\chi]$. We transform this into an interaction picture
\begin{eqnarray}
\tilde \chi(t) &=& e^{i(H_S+H_R)t/\hbar} \chi(t) e^{-i(H_S+H_R)t/\hbar} \\
\tilde H_{SR}(t)&=& e^{i(H_S+H_R)t/\hbar} H_{SR} e^{-i(H_S+H_R)t/\hbar} \quad \\
\dot{\tilde \chi} &=& \frac{1}{i\hbar}[\tilde H_{SR}(t),\tilde \chi(t)] 
\label{von-Neumann}
.
\end{eqnarray}
Note that $\tilde H_{SR}$ picks up an explicit time dependence. We formally integrate the last equation
\begin{eqnarray}
\tilde \chi(t) &=& \tilde \chi(0) + \frac1{i\hbar} \int_0^t dt' [\tilde H_{SR}(t'),\tilde \chi(t')] 
\end{eqnarray}
and substitute this result into the commutator in Eq.\ (\ref{von-Neumann})
\begin{eqnarray}
\dot{\tilde \chi} &=& \frac{1}{i\hbar} [\tilde H_{SR}(t),\tilde \chi(0)]
\nonumber \\ &&
- \frac{1}{\hbar^2} \int_0^t dt' [\tilde H_{SR}(t),[\tilde H_{SR}(t'),\tilde \chi(t')]]
.
\end{eqnarray}
We assume that $R$ is in the vacuum state at $t=0$, uncorrelated with the state of $S$. We assume that $H_{SR}$ is weak and that the reservoir $R$ is much larger than the system $S$, so that the state of $R$ is hardly affected by the interaction with $S$. The correlations between $S$ and $R$ that build up during the time evolution are then weak. In a Born approximation \cite{carmichael:99}, we neglect these correlations and write $\tilde \chi(t)=\tilde \rho(t)R_0$, where $R_0=|0\rangle\langle0|$ is the vacuum density matrix for $R$. We insert this into the right-hand side of the last equation. We then take the partial trace ${\rm Tr}_R$ over the reservoir, assume ${\rm Tr}_R[\tilde H_{SR}(t)R_0]=0$ and obtain \cite{carmichael:99}
\begin{eqnarray}
\label{rho-tilde-dot}
\dot{\tilde \rho}
= - \frac{1}{\hbar^2} \int_0^t dt' {\rm Tr}_R \{[\tilde H_{SR}(t),[\tilde H_{SR}(t'),\tilde \rho(t') R_0]]\}
. \quad
\end{eqnarray}

We write the scattering Hamiltonian (\ref{H-sc}) with the $g_{\alpha'\beta',\alpha\beta}$ from Eqs.\ (\ref{g-zero})--(\ref{g-Hermitian}) and with the assumption $g_{ii,ii}=0$ as
\begin{eqnarray}
H_{SR} &=& \hbar \sum_{{\bf k}} \Gamma^\dag_{\bf k} s_{\bf k} + H.c.
\\
s_{\bf k} &=& \sum_{\bf p}  a_{i,{\bf k}+{\bf p}}  a_{i,{\bf k}-{\bf p}}
\\
\Gamma^\dag_{\bf k} &=& \sum_{\alpha\beta} \sum_{\bf q} \kappa_{\alpha\beta} 
 a_{\alpha,{\bf k}+{\bf q}}^\dag  a_{\beta,{\bf k}-{\bf q}}^\dag
\\
\label{kappa}
\kappa_{\alpha\beta} &=& \frac{g_{\alpha\beta,ii}}{2\hbar L^3}
,
\end{eqnarray}
We recall our assumption that a reservoir state cannot have any particles in state $|i\rangle$. Hence, $s_{{\bf k}}$ acts only on $S$ (internal state $|i\rangle$), whereas $\Gamma_{{\bf k}}$ acts only on $R$ (all other internal states). In addition, our above assumption ${\rm Tr}_R[\tilde H_{SR}(t)R_0]=0$ is satisfied. Moreover, $g_{\alpha'\beta',\alpha\beta}=g_{\beta'\alpha',\beta\alpha}$ because swapping which internal state is labeled first, cannot change the physics. Hence $\kappa_{\alpha\beta}=\kappa_{\beta\alpha}$.

Insertion of $H_{SR}$ into Eq.\ (\ref{rho-tilde-dot}) shows that ${\rm Tr}_R$ vanishes for most of the terms. We obtain
\begin{eqnarray}
\label{rho-tilde-dot-2}
\dot{\tilde \rho}
= - \int_0^t dt' \sum_{\bf k} [\tilde s^\dag_{\bf k}(t), \tilde s_{\bf k}(t') \tilde \rho(t') ]
R_{\bf k}(t,t')
+ H.c.
, \quad
\end{eqnarray}
where we abbreviated $R_{\bf k}(t,t')={\rm Tr}_R \{\tilde \Gamma_{\bf k}(t)\tilde \Gamma^\dag_{\bf k}(t') R_0\}$. We use $\tilde a_{\alpha,{\bf k}}(t)= a_{\alpha,{\bf k}}e^{-i(\omega_\alpha+\hbar k^2 /2m)t}$ and obtain
\begin{eqnarray}
R_{\bf k}(t,t')
&=& \sum_{\alpha\beta} \sum_{\alpha'\beta'} \sum_{{\bf qq}'}
\kappa^*_{\alpha\beta} \kappa_{\alpha'\beta'} 
\nonumber \\ &\times& 
e^{-i(\omega_\alpha+\omega_\beta+\omega_k+\omega_q)t+i(\omega_{\alpha'}+\omega_{\beta'}+\omega_k+\omega_{q'})t'}
\nonumber \\ &\times& 
\langle0| a_{\alpha,{\bf k}+{\bf q}} a_{\beta,{\bf k}-{\bf q}}
a^\dag_{\alpha',{\bf k}+{\bf q}'} a^\dag_{\beta',{\bf k}-{\bf q}'} |0\rangle
,
\qquad 
\end{eqnarray}
For $\alpha\neq\beta$ or ${\bf q}\neq0$ (or both) the last line equals
\begin{eqnarray}
\label{delta}
\delta_{\alpha,\alpha'} \delta_{\beta,\beta'} \delta_{{\bf q},{\bf q}'} 
+ \delta_{\alpha,\beta'} \delta_{\beta,\alpha'} \delta_{{\bf q},-{\bf q}'} 
.
\end{eqnarray}
Using $\kappa_{\alpha\beta}=\kappa_{\beta\alpha}$ we see that both terms produce identical results after carrying out $\sum_{\alpha'\beta'{\bf q}'}$. For $\alpha=\beta$ and ${\bf q}=0$, we obtain only one term, but after carrying out $\sum_{\alpha'\beta'{\bf q}'}$, the result is the same because $\langle 0| a^2 a^{\dag 2} |0\rangle=2$. Hence,
\begin{eqnarray}
R_{\bf k}(t,t')
= 2 \sum_{\alpha\beta} \sum_{{\bf q}} |\kappa_{\alpha\beta}|^2 e^{i(\omega_\alpha+\omega_\beta+\omega_k+\omega_q)(t'-t)}
. \quad
\end{eqnarray}
We now employ a continuum approximation for $\bf q$. At a given center-of-mass momentum, the density of states in ${\bf q}$-space is $(L/2\pi)^3$. To see this, consider first a 1D system. The possible values of the single-particle wave vectors $k_1$ or $k_2$ are separated by $2\pi/L$. Assume that a pair $(k_1,k_2)$ corresponds to a certain center-of-mass momentum. The next possible pair of values that corresponds to the same center-of-mass momentum is $(k_1+L/2\pi,k_2-L/2\pi)$, which changes $q=(k_1-k_2)/2$ by $L/2\pi$. Conversion into 3D and then into frequency space yields the density of states
\begin{eqnarray}
\label{G}
G(\omega_q) = \left( \frac{L}{2\pi} \right)^3 4\pi q^2 \frac{\mu}{\hbar q} 
,
\end{eqnarray}
where $\mu/\hbar q=dq/d\omega_q$. Hence
\begin{eqnarray}
R_{\bf k}(t,t')
&=& 2 \sum_{\alpha\beta} e^{i(\omega_\alpha+\omega_\beta+\omega_k)(t'-t)}
\nonumber \\ && \times 
\int_0^\infty d\omega_q G(\omega_q) |\kappa_{\alpha\beta}|^2 e^{i\omega_q(t'-t)}
. \quad
\end{eqnarray}
$G(\omega_q) |\kappa_{\alpha\beta}|^2$ is typically a slowly varying function of $\omega_q$. For large $t-t'$, the phase factor $e^{i\omega_q(t'-t)}$ oscillates rapidly and thus the integral is almost zero. This is the rapid dephasing of the reservoir stats that we mentioned earlier. The integral over $t'$ in Eq.\ (\ref{rho-tilde-dot-2}) is thus dominated by the values at $t'\approx t$. This justifies a Markov approximation \cite{carmichael:99}, which consists of replacing $\tilde \rho(t')$ in Eq.\ (\ref{rho-tilde-dot-2}) by $\tilde \rho(t)$. Since the integral over $t'$ is dominated by the values at $t'\approx t$, we can extend the lower bound of the integral to $t'=-\infty$ \cite{carmichael:99}. We substitute $\tau=t-t'$ and Eq.\ (\ref{rho-tilde-dot-2}) becomes
\begin{eqnarray}
\label{rho-tilde-dot-3}
\dot{\tilde \rho}
&=& - \sum_{{\bf kpp}'} [ 
a^\dag_{i,{\bf k}+{\bf p}} a^\dag_{i,{\bf k}-{\bf p}}
, a_{i,{\bf k}+{\bf p}'} a_{i,{\bf k}-{\bf p}'} \tilde \rho(t) ] B_{{\bf kpp}'}(t)
\nonumber \\ && 
+ H.c.
\end{eqnarray}
with
\begin{eqnarray}
B_{{\bf kpp}'}(t) 
&=& \int_0^\infty d\tau  R_{\bf k}(t,t') e^{i(\omega_k+\omega_p)t-i(\omega_k+\omega_{p'})t'}
\qquad
\\
&=& 2 e^{i(\omega_p-\omega_{p'})t} \sum_{\alpha\beta} \int_0^\infty d\tau \int_0^\infty d\omega_q 
\nonumber \\ && \times
G(\omega_q) |\kappa_{\alpha\beta}|^2 e^{-i(\omega_\alpha+\omega_\beta+\omega_q-\omega_{p'})\tau}
.
\end{eqnarray}
We swap the two integrals, performing the time integration first. We use \cite{carmichael:99}
\begin{eqnarray}
\int_0^\infty d\tau e^{-i\omega\tau} = \pi \delta(\omega) -i \frac{P}{\omega}
,
\end{eqnarray}
where $P$ denotes the principal value for the following $d\omega$ integration. Hence
\begin{eqnarray}
B_{{\bf kpp}'}(t) 
&=& e^{i(\omega_p-\omega_{p'})t} \Big( \frac{\gamma_D}{2} + i \Delta \Big)
\\
\label{gamma-D}
\frac{\gamma_D}{2}
&=& 2\pi \sum_{\alpha\beta} G(\omega_q) |\kappa_{\alpha\beta}|^2
\\
\label{Delta-P}
\Delta
&=& - 2\sum_{\alpha\beta} P \int_0^\infty \frac{G(\omega_q) |\kappa_{\alpha\beta}|^2 d\omega_q} {\omega_q-(\omega_{p'}-\omega_\alpha-\omega_\beta)}
, \qquad
\end{eqnarray}
where $G(\omega_q) |\kappa_{\alpha\beta}|^2$ in Eq.\ (\ref{gamma-D}) is to be taken at $\omega_q=\omega_{p'}-\omega_\alpha-\omega_\beta$, which reflects energy conservation.

Note that our results for $\gamma_D$ and $\Delta$ are twice as large as Eqs.\ (1.68) and (1.70) in Ref.\ \cite{carmichael:99}. This difference is due to the two terms in Eq.\ (\ref{delta}). Ref.\ \cite{carmichael:99} considers reservoir states with only one particle, so that only one term occurs in the expression analogous to Eq.\ (\ref{delta}).

We transform Eq.\ (\ref{rho-tilde-dot-3}) back from the interaction picture to the Schr\"odinger picture. We use that $\gamma_D$ and $\Delta$ are real and obtain
\begin{eqnarray}
\label{rho-tilde-dot-4}
\lefteqn{
\dot \rho
+ \frac{i}{\hbar} [H_S+H_\Delta,\rho]
= }
\\ \nonumber 
&=& 
\sum_{{\bf kpp}'} \frac{\gamma_D}2
\left[ a_{i,{\bf k}+{\bf p}'} a_{i,{\bf k}-{\bf p'}} \rho(t), a_{i,{\bf k}+{\bf p}}^\dag a_{i,{\bf k}-{\bf p}}^\dag \right] 
+H.c.
\end{eqnarray}
with
\begin{eqnarray}
\label{H-Delta}
H_\Delta
= \sum_{{\bf kpp}'} (\hbar \Delta )
a_{i,{\bf k}+{\bf p}}^\dag a_{i,{\bf k}-{\bf p}}^\dag a_{i,{\bf k}+{\bf p}'} a_{i,{\bf k}-{\bf p'}} 
.
\end{eqnarray}
$H_\Delta$ is very reminiscent of the energy shift in ordinary second-order perturbation theory, because $H_\Delta$ describes processes in which a pair of particles in state $|ii\rangle$ makes a transition to a state $|\alpha\beta\rangle$ with relative momentum $\hbar{\bf q}$ and then returns to state $|ii\rangle$. Energy is usually not conserved in such virtual transitions and the energy difference between the initial and intermediate state appears in the denominator of Eq.\ (\ref{Delta-P}). This denominator gives the typical weighting for such virtual transitions, known from perturbation theory.

A calculation of the value of $\Delta$ from Eq.\ (\ref{Delta-P}) is difficult, even if the incoming particles are ultracold. This is because, first, Eq.\ (\ref{sum-g-fi}) cannot be used because the energy denominator depends on $\alpha$ and $\beta$ and, second, because energy is not conserved in the virtual transitions so that a calculation of the integral requires knowledge of the $S$-matrix elements in Eq.\ (\ref{g-alpha-beta-ii}) over a large range of energies $\hbar\omega_q$, even if $\hbar\omega_{p'}$ is fixed to be ultracold. The energy dependence of the $S$-matrix elements is sensitive to many details of the true potential. A calculation of $\Delta$ is thus beyond the scope of the present work.

We see that $H_\Delta$ in Eq.\ (\ref{H-Delta}) has exactly the same form as the elastic collision term $(\alpha,\beta)=(\alpha',\beta')=(i,i)$ in Eq.\ (\ref{H-sc}). In the ultracold limit, $H_\Delta$ thus simply contributes to the mean-field energy. Any attempt to measure ${\rm Re}(a)$ in the many-body system will always yield the sum of the two contributions, so that we can simply absorb $\Delta$ in an effective value of ${\rm Re}(a)$, that needs to be determined experimentally.

A calculation of $\gamma_D$ from Eq.\ (\ref{gamma-D}) is much simpler because it requires knowledge of the $S$-matrix elements only for outgoing momenta $\hbar{\bf q}$ that satisfy energy conservation. There is no additional denominator and we can use Eqs.\ (\ref{sum-g-fi}), (\ref{kappa}) and (\ref{G}) with ${\bf q}={\bf k}_{\alpha\beta}$ and ${\bf p}'={\bf k}_{ii}$. We obtain
\begin{eqnarray}
\gamma_D = \frac{4\pi\hbar}{L^3 m} \ \frac{{\rm Im}(\tan\eta)}{k_{ii}}
.
\end{eqnarray}
For ultracold incoming particles, we use Eqs.\ (\ref{a}) and (\ref{g-3D}) and obtain
\begin{eqnarray}
\gamma_D = - \frac{{\rm Im}(g_{\rm 3D})}{\hbar L^3}
.
\end{eqnarray}
This is independent of ${\bf k}$, ${\bf p}$, and ${\bf q}$, so that Eq.\ (\ref{rho-tilde-dot-4}) can be transformed into a simple expression in position representation. This yields Eq.\ (\ref{D}), which is the central result of this appendix.

\subsection{Comparison with Existing Literature}
\label{app-scattering}
Finally, we discuss how our results relate to the existing literature on scattering theory. To this end, we note that the rate coefficient $K_{\rm 3D}$ in Eq.\ (\ref{K-3D}) can be calculated in an alternative way, based on the observation that the scattering rate per particle in a gas is proportional to $n_{\rm 3D} \sigma \hbar k_{\rm rel}/m$, where $\sigma$ is the cross section and ${\bf k}_{\rm rel}$ is the wave vector of the relative motion of the two particles before the collision. For elastic and inelastic two-body collisions of identical bosons one obtains \cite{bohn:97,balakrishnan:97}
\begin{eqnarray}
\sigma_{\rm el} &=& 8\pi|a|^2 \\
\sigma_{\rm inel} &=& - \; \frac{8\pi}{k_{\rm rel}} {\rm Im}(a)
,
\end{eqnarray}
respectively. Deriving the rate coefficient $K_{\rm 3D}$ from $\sigma_{\rm inel}$ is a delicate issue because of several subtle factors of 2. An unambiguous way to handle those factors is based on a quantum Boltzmann equation \cite{stoof:89} and confirms our result Eq.\ (\ref{K-3D}).

\section{Applicability of the Model to the Experiment in Ref.\ \cite{syassen:08}}
\label{app-experiment}
This appendix discusses the applicability of the model presented in Sec.\ \ref{sec-loss-1D} to our experiment reported in Ref.\ \cite{syassen:08}. First, we note that the derivation of $g_{\rm 1D}$ in Ref.\ \cite{olshanii:98} relies on the assumption that the kinetic energy in the relative motion of two particles $\hbar^2 k_{\rm rel}^2/m$ before a collisions is less than the splitting of the transverse harmonic oscillator levels with even angular momentum $2\hbar\omega_\perp$. This is equivalent to $k_{\rm rel} a_\perp < \sqrt2$.

In a fermionized ground state, noticeable momenta can occur even at zero temperature. We can use the $k_j$ to estimate the typical momenta in the ground state. We note that the $k_j$ are not actually momenta, see {\it e.g.}\ Ref.\ \cite{girardeau:01}, but the typical width of the momentum distribution is set by $k_j$. We are interested in the typical relative momentum in the fermionized ground state, which we can thus estimate from Eq.\ (\ref{k-j}) as $k_{\rm rel} \lesssim k_N-k_1=(N-1)2\pi/L\approx 2\pi n_{\rm 1D}$. The experiment in Ref.\ \cite{syassen:08} was performed at an initial density of $n_{\rm 1D}=2/(830\ {\rm nm})$ and with $a_\perp=39$ nm, so that $k_{\rm rel} a_\perp \lesssim 0.60$ which is well inside the regime $k_{\rm rel} a_\perp < \sqrt2$.

Second, we mention another assumption in the derivation of $g_{\rm 1D}$ in Ref.\ \cite{olshanii:98}, namely that the 1D scattering process can be modeled by a 1D delta potential only for low enough momenta. For $k_{\rm rel} a_\perp \leq 0.60$, ${\rm Re}(a)=0$, and $K_{3D}=1.5\times 10^{-10}$ cm$^3$/s \cite{syassen:06} we numerically evaluate the analytic results of Ref.\ \cite{olshanii:98}. We find that the transmission coefficient \cite{olshanii:98} for the scattering process is altered by less than 10\% when making the transition from the 3D regularized delta potential to the 1D delta potential, thus showing that the model is, indeed, a good approximation.

Finally, we briefly discuss the part of our experiment in Ref.\ \cite{syassen:08} in which a weak lattice is applied along the one dimension. If two particles occupy the same lattice site, the resulting on-site interaction is described by a matrix element $U$. We will discuss now, why the expression $U = g_{\rm 3D} \int d^3x \; |w({\bf x})|^4$ is valid, where $w({\bf x})$ is a Wannier function. First, the lattice potential at a single lattice site is to a good approximation harmonic, with an aspect ratio of less than 3 for the parameters of our experiment. Scattering processes within one lattice site are therefore not in a 1D regime and only for much larger aspect ratios would we obtain $U = g_{\rm 1D} \int dx \; |w(x)|^4$, with a 1D Wannier function. Second, if two particles occupy the same site, this 3D system at one site has $|a|^3 n_{3D}\ll 1$. To see this we consider a lattice depth of $V_0=127 E_r$ \cite{syassen:08} along all three dimensions, where $E_r$ is the recoil energy, yielding a harmonic oscillator length of $a_{ho}=39$ nm at one site. At $K_{3D}=1.5\times 10^{-10}$ cm$^3$/s \cite{syassen:06} and ${\rm Re}(a)=0$, we obtain for two particles at one site $|a|^3 n_{3D,\rm peak}=2 |a|^3/(\sqrt{\pi} a_{ho})^3=0.03$. The fact that $|a|^3 n_{3D}\ll 1$ means that two particles at an isolated single site are in the weakly correlated regime and the single-particle Wannier function can be used to calculate~$U$.

\acknowledgments

We thank Paul Julienne for enlightening discussions about the subtle factors of 2 in appendix \ref{app-scattering}. We acknowledge financial support of the German Excellence Initiative via the  program Nanosystems Initiative Munich and of the Deutsche Forschungsgemeinschaft via SFB 631. JJGR acknowledges financial support from the Ramon y Cajal Program and the Spanish projects FIS2006-04885 and CAM-UCM/910758.

\end{document}